\documentclass[twocolumn,english]{IEEEtran}
\usepackage[T1]{fontenc}
\usepackage[latin9]{inputenc}
\usepackage{color}
\usepackage{units}
\usepackage{amsmath}
\usepackage{amssymb}
\usepackage{graphicx}
\usepackage[numbers,sort&compress]{natbib}

\makeatletter
\usepackage[figurename=Fig.]{caption}

\makeatother

\usepackage{babel}
\begin{document}
\title{Ultra-Sensitive Radio Frequency Biosensor at an Exceptional Point
of Degeneracy induced by Time Modulation}
\author{\author{Hamidreza Kazemi, Amirhossein Hajiaghajani, Mohamed Y. Nada, Manik Dautta, Muhannad Alshetaiwi, Peter Tseng, and Filippo Capolino\thanks{H. Kazemi, A. Hajiaghajani, M. Y. Nada, M. Dautta, M. Alshetaiwi, P. Tseng, and F. Capolino are with Department of Electrical Engineering and Computer Science, University of California, Irvine, CA 92697, USA (e-mail: \{hkazemiv,ahajiagh,mynada,mdautta,malsheta,tsengpc,f.capolino\}@uci.edu).}
\thanks{P. Tseng is also with the Department of Biomedical Engineering, University of California, Irvine, CA 92697, USA.}}}
\maketitle
\begin{abstract}
We exploit the premises of exceptional points of degeneracy (EPDs)
induced in linear time-periodic (LTP) systems to achieve extremely
sensitive biosensors. The EPD is formed in a single LC resonator where
the total capacitance is comprised of a time-varying capacitor in
parallel to a biosensing capacitor. We use the time-periodic variation
of a system parameter (e.g., capacitance) to achieve a second order
EPD aiming at improving the sensitivity of liquid based radio frequency
biosensors, leading to an intrinsic ultra sensitivity. We show the
emergence of EPDs in such a system and the ultra sensitivity of the
degenerate resonance frequency to perturbations compared to conventional
RF sensors. Moreover, we investigate the capacitance and conductance
variations of an interdigitated biosensing capacitor to the changes
in the concentration of a biological material under test (MUT), leading
to subsequent large changes in the resonance frequency of the LTP-LC
resonator. A comparison with a standard LC resonator demonstrates
the ultra-high sensitivity of the proposed LTP-LC based biosensor.
In addition, we show the scalability of the biosensor sensitivity
across different frequency ranges.
\end{abstract}

\section{Introduction}

\IEEEPARstart{A}{nalytical} biosensors play a tremendous role in
modern medicine through enabling the monitoring of biomarkers in human
health. The applications of these sensors are diverse as they form
the core of many point-of-care, wearable, and diagnostic tools utilized
in pathology, nutrition, fitness, biomedical science, and more \citep{wang2006electrochemical,wei2016microfluidic,mandl2009endoplasmic,Lopez2010LOBIN,Huang2009Prevasive}.
Traditionally, an analytical biosensor is composed of two main elements:
a bioamplifier (such as a bioreceptor), and a transducer (converting
the biological signal into an electrical one). Various number of modalities
exist to monitor biomarkers \citep{kim2019wearable,lee2018trends},
including but not limited to electrochemical impedance spectroscopy
\citep{EIS,EIS_1,EIS_2}, piezoelectric microcantilever \citep{loo2011rapid,loo2011highly,capobianco2008label},
surface plasmon resonance \citep{MEYER,erturk2016microcontact,guo2016highly,sun2017aptasensors},
immunoelectrophoresis \citep{yman2006food,Immunoelectrophoresis},
fluorescence \citep{wang2015fluorescent,Fluroscent}, enzyme-linked
immunosorbent assay (ELISA) \citep{ELISA,Rica2012,jia2009,holzhauser2002}.
While many of these techniques have found critical roles in a variety
of applications, a majority are encumbered by limitations in system
size and weight, sample preparation requirements, power consumption,
and limited capabilities in wireless operation.

Dielectric-RF sensors (that sense the presence of analytes via permittivity
shifts) possess traits that address many issues that have limited
traditional biosensors, however these sensors have limited use in
modern devices. The reason for this is two-fold: these sensors possess
low sensitivity (signal change due to input) and low selectivity (discrimination
of an analyte from interferents). Moreover, RF biosensors having various
capabilities are attractive among other sensing methods and gained
a lot of attention since their working principle is dependent on the
change of dielectric properties of a medium which is a label free
fast detection method compare to other methods such as electrochemical
or piezoelectric sensors which requires complex wiring. While there
are potential strategies to improve RF-biosensor selectivity \citep{tseng2018functional,dautta2020passive,kim2015rapid,hajiaghajani2019selective},
here we examine a new electromagnetic amplification strategy to improve
biosensor sensitivity. \captionsetup[figure]{labelformat={default},labelsep=period,name={Fig.}}

\begin{figure}
\begin{centering}
\includegraphics[width=3.5in]{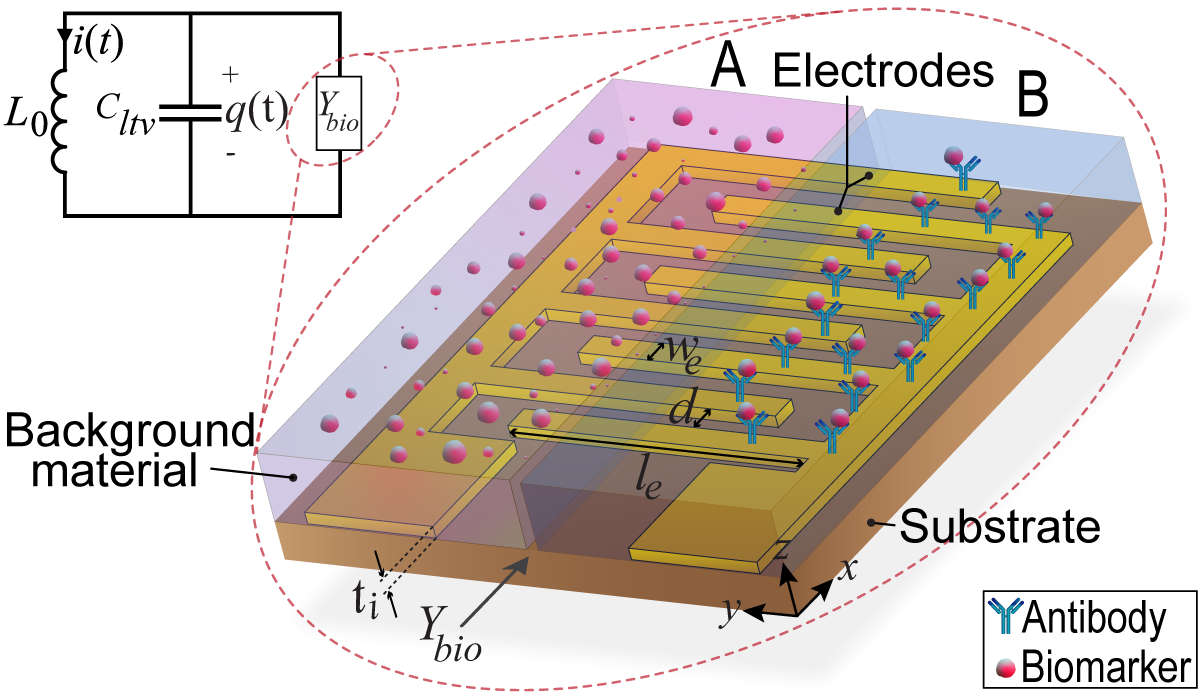}
\par\end{centering}
\caption{\label{fig:BioSensor}The proposed sensor circuit working at an exceptional
point of degeneracy consisting of a time-varying LC resonator in parallel
to a biosensing capacitor (its capacitance is function of the concentration
of the MUT). The biosensing capacitor is realized using an interdigitated
capacitor. The EPD induced in this single LTP-LC resonator is responsible
of the very high sensitivity.}
\end{figure}
One of the most recent methods to dramatically enhance sensors sensitivity
is to design the RF sensor to operate at the so called exceptional
point of degeneracy. The EPD represents the coalescing point of the
degenerate resonance frequencies and it emerges in a system when two
or more eigenmodes of the system coalesce into a single degenerate
eigenmode in both their eigenvalues and eigenvectors \citep{bender_real_1998,figotin_oblique_2003,guo_observation_2009,othman_low_2016,othman_exceptional_2017,veysi_degenerate_2018,nada2018various}.
The system at an EPD shows an inherent ultra sensitivity, specially
for small perturbation, that can be exploited to enhance the sensitivity
of the liquid-based radio frequency biosensor \citep{Kazemi2019Exceptional,wiersig_enhancing_2014,wiersig2016sensors,hodaei_enhanced_2017,nada2018microwave,abdelshafy2019exceptional}.
In addition to high sensitivity, EPDs are associated with other unique
properties such as enhancing the gain of active systems \citep{othman_giant_2016-1},
lowering oscillation threshold \citep{oshmarin2019new}, etc.

Recently, EPD-induced sensitivity based on the concept
of parity-time (PT) symmetry in multiple, coupled resonators has been
investigated \citep{Sakhdari2018ultrasensitive,Chen2018Generalized}.
EPDs realized in PT-symmetric systems would require at least two coupled
resonators, and precise knowledge of gain and loss in the system.
In contrast, in this paper we employ EPDs directly induced via time
modulation of a component in a \textit{single}
resonator \citep{Kazemi2019Exceptional,Kazemi2019Expre} to conceive
a new class of biosensors. EPDs induced by time modulation-only require
a single resonator and are easily tuned by changing the modulation
frequency of a component which is a viable strategy to obtain EPDs
since varying frequency in a precise manner is common practice in
electronic systems. In this paper, we use the concept of an EPD that
occurs in a single resonator and it is not based on PT-symmetry
\citep{Kazemi2019Exceptional}; this new method is used to generate
a second order EPD induced by time-periodic variation of a system
parameter \citep{Kazemi2019Exceptional} aiming at improving the sensitivity
of liquid based radio frequency biosensors, leading to an intrinsic
ultra sensitivity. The concept of an EPD in a single resonator obtained
by simply applying a time domain modulation was shown in \citep{Kazemi2019Exceptional}
and the experimental demonstration of the occurrence of such EPD has
been shown in \citep{Kazemi2019Expre}. The proposed biosensor shown
in Fig. \ref{fig:BioSensor} is comprised of an LC resonator where
the capacitor is time-variant and is in parallel to the biosensing
capacitor, i.e., the capacitor whose capacitance is function of the
concentration of the MUT. The biosensing capacitor is implemented
using an interdigitated capacitor (IDC) as shown in Fig. \ref{fig:BioSensor}.
In this system, the change in the concentration of the MUT will change
the capacitance of the IDC that can be measured through the shift
in the resonance frequency of the system and this shift is boosted
when the system operates at an EPD. We study two different biosensing
scenarios based on the IDC in Fig. \ref{fig:BioSensor}: (i) a uniformly
dissolved MUT in the background material above the IDC, and (ii) a
thin layer of MUT placed on top of the electrodes which are denoted
by A and B in Fig. \ref{fig:BioSensor}. Note that these two scenarios
are combined into one figure for brevity, however, for the analysis
we consider each scenario separately.

In the following, we first show the behavior of a LTP-LC resonator
through the dispersion relation of the resonance frequency versus
modulation and we discuss the occurrence of EPDs in such a system.
The analysis accounts for losses in the system. In section \ref{sec:Analysis-of-Bio}
we design and investigate the performance of an IDC which is integrated
in the system as the biosensing capacitor. We show the effect of the
concentration of the MUT on the capacitance and the conductance of
such capacitor for two cases of uniformly dissolved MUT and effective
MUT layer. Finally, in section \ref{sec:Characterization-of-the},
we show the sensitivity of the designed system to perturbation, i.e.,
the concentration of the MUT, and we characterize the proposed biosensor
performance across different designs and frequencies. Moreover, to
show the advantages and the superiority of the proposed EPD biosensor,
we compare its sensitivity with that of conventional biosensors.

\section{Enhancing the sensitivity of biosensors in an LTP system with EPDs\label{sec:Enhance_sens}}

\begin{figure}
\begin{centering}
\includegraphics[width=3.5in]{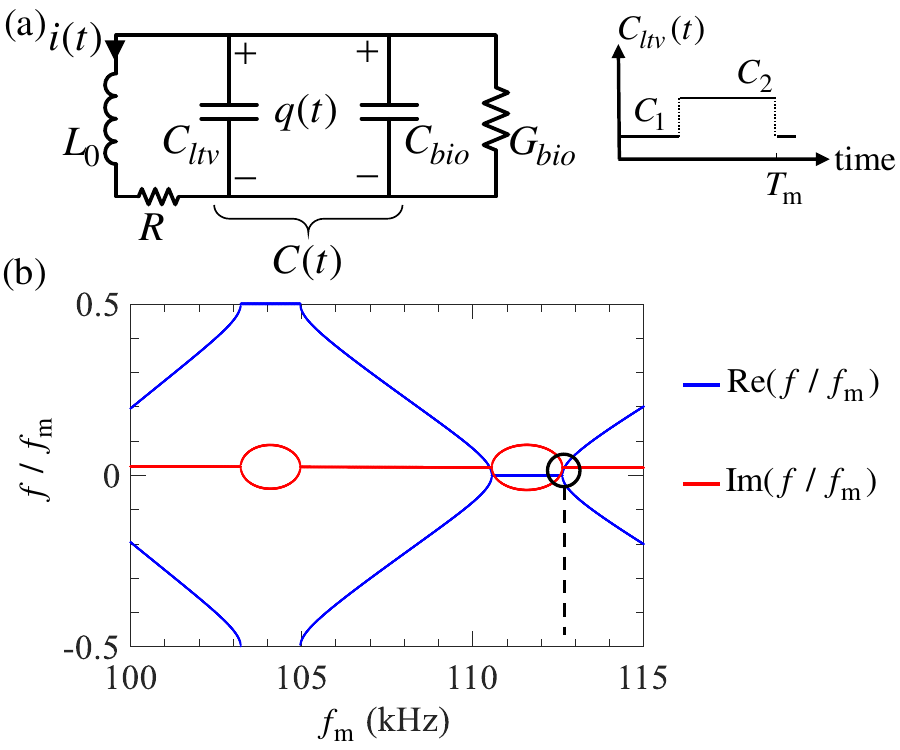}
\par\end{centering}
\caption{\label{fig:LC_circuit}(a) The proposed LTP-LC resonator sensing circuit
consisting of and inductor $L_{0}$ in series with a resistance $R$
, in parallel to a time-varying capacitor $C_{ltv}$ whose capacitance
is given by a two level piece-wise constant time-periodic function
as shown in the subset, with modulation frequency $f_{\mathrm{m}}$.
The LTP-LC tank also includes a parallel biosensor capacitor $C_{bio}$
and associated conductance $G_{bio}$. The values of both $C_{bio}$
and $G_{bio}$ are function of the concentration $\delta$ of the
MUT. (b) Dispersion diagram of the circuit resonance frequencies vs.
modulation frequency $f_{\mathrm{m}}$ of $C_{ltv}$. The blue and
red curves show the real and imaginary parts of the resonance frequency,
respectively. The dispersion diagram accounts for small losses in
the circuit.}
\end{figure}

In this section, we demonstrate how to boost the sensitivity of conventional
biosensors using a LTP-LC resonator as indicated in Fig. \ref{fig:LC_circuit}(a)
where the time-periodic variation is introduced in the system through
the time-varying capacitor, $C_{ltv}(t)$. The two-dimensional state
vector $\boldsymbol{\Psi}(t)=[q(t),i(t)]^{T}$ describes this system,
where $T$ denotes the transpose operator, $q(t)$ and $i(t)$ are
the capacitor charge on both the capacitors in Fig. 2(a) and inductor
current, respectively. The temporal evolution of the state vector
obeys the two-dimensional first-order differential equation \citep{Kazemi2019Exceptional}

\begin{equation}
\frac{\mathrm{d}}{\mathrm{d}t}\mathbf{\Psi}(t)=\underline{\mathbf{M}}(t)\mathbf{\Psi}(t)
\end{equation}
where $\underline{\mathbf{M}}(t)$ is the $2\times2$ time-varying
system matrix. Assuming that the time-variation of the capacitance
is a two level piece-wise constant, periodic, function as shown in
the subset of Fig. \ref{fig:LC_circuit}(a), the time-variant system
matrix reads

\begin{equation}
\underline{\mathbf{M}}_{p}=\left[\begin{array}{cc}
-\nicefrac{G_{bio}}{(C_{p}+C_{bio})} & -1\\
\nicefrac{1}{(L_{0}(C_{p}+C_{bio}))} & -\nicefrac{R}{L_{0}}
\end{array}\right],
\end{equation}
where $C_{p}$, with $p=1,2,$ represents the two values of the piece-wise
constant time-varying capacitance $C_{ltv}(t)$ and $C_{bio}$ is
the capacitance of the biosensing capacitor. The linear time-varying
capacitance $C_{ltv}(t)$ is $C_{ltv}=C_{1}$ for $0<t\leq0.5T_{\mathrm{m}}$
and $C_{ltv}=C_{2}$ for $0.5T_{\mathrm{m}}<t\leq T_{\mathrm{m}}$.
Losses in the system are represented by the series resistance of the
inductor $R$ and the parallel conductance $G_{bio}$ of the biosensing
capacitor. The conductance $G_{bio}$ represents losses in the background
medium and in the MUT.

Considering that the LTP sensor is periodic with period $T_{\mathrm{m}}=1/f_{\mathrm{m}}$,
we can translate the state vector from the time instant $t$ to $t+T_{\mathrm{m}}$
as $\mathbf{\Psi}(t+T_{\mathrm{m}})=\underline{\mathbf{\Phi}}(t,t+T_{\mathrm{m}})\mathbf{\Psi}(t)$
through the $2\times2$ state transition matrix $\underline{\mathbf{\Phi}}(t,t+T_{\mathrm{m}})$
\citep{Kazemi2019Exceptional,richards2012analysis}. In addition,
the state vector satisfies $\mathbf{\Psi}(t+T_{\mathrm{m}})=e^{\mathrm{j}\omega T_{\mathrm{m}}}\mathbf{\Psi}(t)$
in a periodic systems, hence we constitute the eigenvalue problem
as

\begin{equation}
\left(\underline{\mathbf{\Phi}}(t,t+T_{\mathrm{m}})-e^{\mathrm{j}\omega T_{\mathrm{m}}}\underline{\mathbf{I}}\right)\mathbf{\Psi}(t)=0,\label{eq:eig_prob}
\end{equation}
where $\underline{\mathbf{I}}$ is the $2\times2$ identity matrix.
Considering the eigenvalue problem derived in (\ref{eq:eig_prob}),
we find the eigenvalues $e^{\mathrm{j}\omega T_{\mathrm{m}}}$ of
the state transition matrix $\underline{\mathbf{\Phi}}(t,t+T_{\mathrm{m}})$,
hence the circuit eigenfrequencies $f=\omega/(2\pi)$ that are the
resonance frequencies of the circuit. Figure \ref{fig:LC_circuit}(b)
shows the dispersion of these LTP-LC resonant frequencies versus modulation
frequency $f_{\mathrm{m}}$. The small asymmetry of the real and imaginary
parts of the resonance frequencies $f$ with respect to the center
$f=0$ is due to the small losses in the circuit components. Such
a dispersion diagram is obtained for the circuit parameters set as
$L_{0}=15\mathrm{\mu H}$, $R=0.1\Omega$, $C_{1}=4.5\,\mathrm{nF}$,
$C_{2}=1.5\,\mathrm{nF}$. The parameters of the biosensing capacitor
are derived based on the first order model described in Section \ref{sec:Analysis-of-Bio}
and set as $C_{bio}=0.3\,\mathrm{nF}$ and $G_{bio}=67\,\mathrm{\mu S}$.
It is observed from Fig. \ref{fig:LC_circuit}(b) that the time-periodic
LC resonator exhibits second order EPDs (the band edges of each band
gap) for selected modulation frequencies, i.e., when two resonance
frequencies coalesce at a specific modulation frequency $f_{\mathrm{m}}$.
Note that the LC resonator is time-periodic, therefore for a resonance
frequency $f$ there exist all the correspondent Fourier harmonics
$f+nf_{\mathrm{m}}$, where $n=\pm1,\pm2,...$. The EPDs occur either
at the center or at edge of the Brillouin zone (BZ) (we use this term
in analogy to what happens in periodic electromagnetic waveguides
\citep{abdelshafy2019exceptional}) as it can be inferred from Fig.
\ref{fig:LC_circuit}(b). For instance, one of the EPDs in Fig. \ref{fig:LC_circuit}(b)
is indicated with a black circle: at the modulation frequency $f_{\mathrm{m}}=112.6\,\mathrm{kHz}$
the LTP-LC resonance frequencies are $f_{e}=f_{e0}+nf_{\mathrm{m}}$.
The one in the black circle corresponds to the $n=0$ harmonic $f_{e0}=\mathrm{j2.6\,kHz}$,
and the small imaginary part is due to losses in the circuit.

The resonance frequencies of such a system operating at an EPD are
highly sensitive to perturbation of any system parameter. In general,
a perturbation $\delta$ of a system parameter leads to a perturbed
transition matrix $\underline{\boldsymbol{\Phi}}(\delta)$ that in
turn generates two perturbed resonant frequencies $f_{p}(\delta)$,
with $p=1,2$, slightly away from the degenerate resonance frequency
$f_{e}$ of the system operating at the second order EPD. It has been
demonstrated that the perturbation of the eigenvalues of (\ref{eq:eig_prob}),
hence the perturbation of the resonant frequencies, cannot be represented
with a Taylor expansion of the degenerate resonant frequency around
$f_{e}$ \citep[chapter II.1.1]{kato_perturbation_1995}. The first
order approximation of $f_{p}(\delta)$ near the EPD is derived by
a Puiseux series \citep[chapter II.1.1]{kato_perturbation_1995} (also
called ``fractional power expansion'') using the explicit recursive
formulas given in \citep{welters_explicit_2011,Kazemi2019Expre} as

\begin{equation}
f_{p}(\delta)\approx f_{e}\pm\mathrm{j}\frac{f_{\mathrm{m}}}{2\pi}(-1)^{\mathit{p}}\alpha_{1}\sqrt{\delta}\label{eq:f_p}
\end{equation}
where $\alpha_{1}=\sqrt{-\frac{\mathrm{d}}{\mathrm{d}\delta}\left[\mathrm{det}(\boldsymbol{\underline{\Phi}}(\delta)-e^{\mathrm{j}2\pi fT_{\mathrm{m}}}\underline{\mathbf{I}})\right]}\mid_{\delta=0,f=f_{e}}$
and the $\pm$ signs correspond to EPDs at the center or edge of the
BZ. Equation (\ref{eq:f_p}) indicates that for a small perturbation
$\delta<<1$, the resonance frequencies $f_{p}$ change dramatically
from their original degenerate frequency $f_{e}$ due to the square
root function. As an example, a perturbation $\delta=0.0001$ generates
a resonance frequency shift $f_{p}(\delta)-f_{e}$ proportional to
$\sqrt{\delta}=0.01$, that is much larger than that in standard LC
resonators, where such shift would be simply proportional to $\delta$.

Note that in our proposed scheme with a single LTP
resonator, perturbation of a system parameter (typically the sensing
capacitance) perturbs the system away from the EPD and this results
in two shifted resonances $f_{p}(\delta)$ from the EPD frequency
$f_{e}$. The two shifts $f_{p}(\delta)-f_{e}$ are real valued and,
as discussed above, a small imaginary part in the EPD frequency is
present because of losses. This is in contrast to what occurs in a
two-coupled resonator system operating at an EPD based on PT symmetry:
indeed, in the PT symmetric system, perturbing the sensing capacitor
on the lossy (sensing) side only, disqualifies the system as being
PT symmetric and as a result both of the resonant frequencies become
complex. Therefore, in order to maintain the PT symmetry and exceptional
sensitivity, the prior knowledge of the sensing capacitance change
is required to vary also the capacitance on the active part of the
PT symmetry circuit. Finally, note that depending on the choice of
the operating EPD, in our proposed scheme one may have purely real
or purely imaginary resonant frequency shifts, depending on the sign
of $\delta$, which will be discussed in more details in Sec. \ref{sec:Characterization-of-the}.

\section{Analysis of biosensing capacitor\label{sec:Analysis-of-Bio}}

An IDC is considered here as a reliable candidate to realize the biosensing
capacitor $C_{bio}$. The capacitance and conductance of the IDC depend
on the geometrical design and the electromagnetic properties of the
various materials used to fabricate the capacitor. We propose an equivalent
circuit model for the IDC that easily describes the sensor's port
admittance as shown in Fig. \ref{fig:Unprtb and C}, and that will
be used for the perturbation analysis in the next subsection.

In order to simplify the analysis of the IDC and obtain an equivalent
circuit model, we divide the media surrounding the electrodes into
several domains based on their electromagnetic properties and electric
field distribution. Hence, we model each domain by a parallel capacitance
and conductance, placed along the direction of the electric field.
The cross section of the biosensing capacitor showing the different
domains is illustrated in Fig. \ref{fig:Unprtb and C}(a) where only
two electrodes of the interdigitated capacitor are shown. The domains
are the substrate, the inter-electrode layer, and the background material
above the electrodes. The equivalent circuit models of such domains
are in parallel (see Fig. \ref{fig:Unprtb and C}(a)). In practice,
the background and the substrate domains are significantly thicker
than the inter-electrode gap, hence they are approximated by two infinite
half-spaces. The fringing electric field distribution between two
electrodes of the IDC cell is similar to that of a pair of parallel
strip lines. We derive analytical formulas for capacitances and conductances
of the equivalent circuit model of the cell shown in Fig. \ref{fig:Unprtb and C}(a)
by fitting the result of the admittance formula of two parallel strip
lines to that of numerical simulations. The numerical results are
obtained by finite element method (FEM) simulations implemented in
COMSOL Multiphysics$\circledR$ \citep{Comsol}. Typically, the geometrical
ratios in strip lines are different from IDC, and to derive accurate
ad-hoc formulas, we add correction factors in the analytic expressions
for the capacitance and conductance of the parallel strip lines (whose
approximation is available in \citep{garg2013microstrip}):

\begin{figure}
\begin{raggedright}
(a)~~~~
\par\end{raggedright}
\begin{centering}
\includegraphics[width=2.5in]{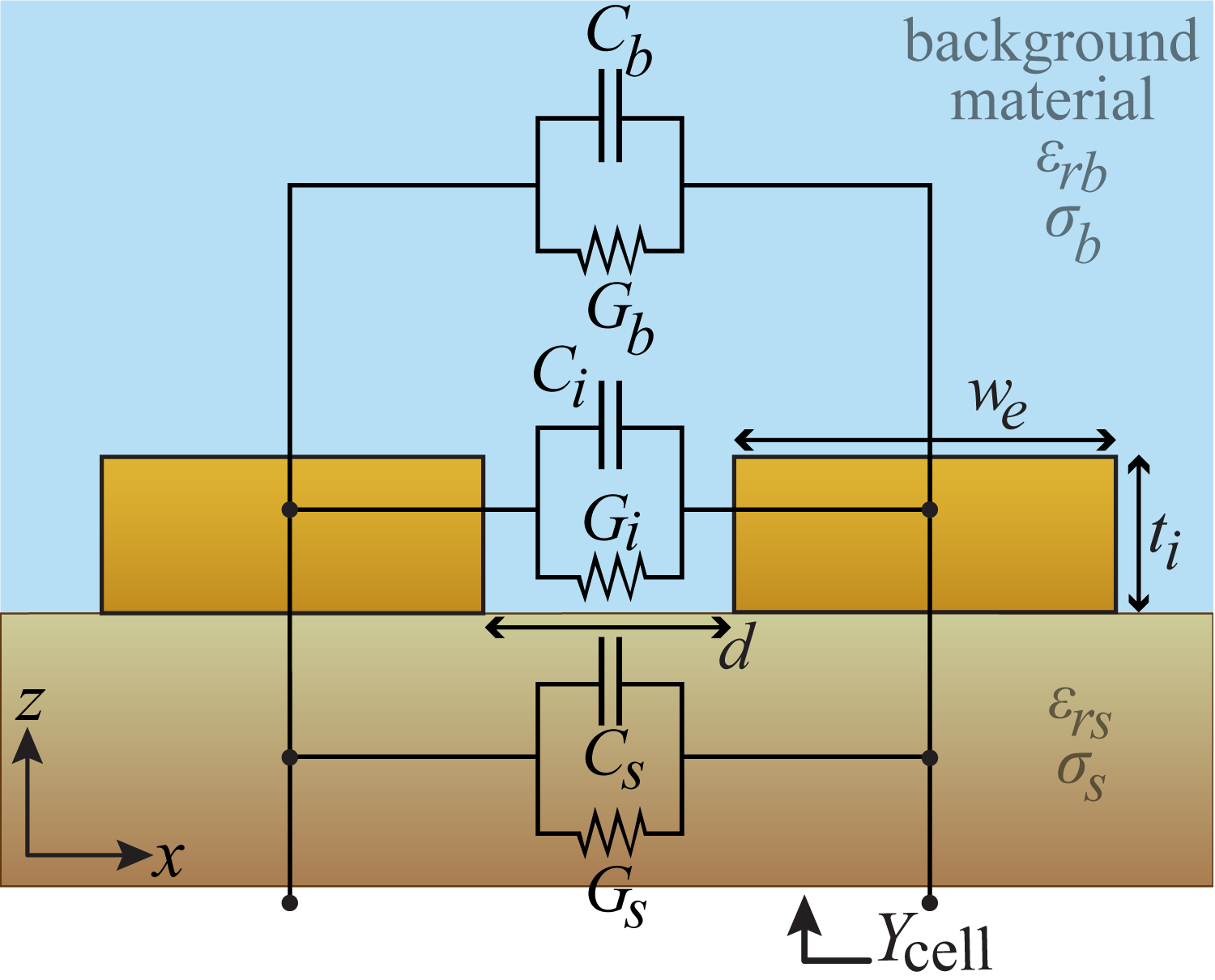}
\par\end{centering}
\begin{raggedright}
(b)
\par\end{raggedright}
\begin{centering}
\includegraphics[width=2.1in]{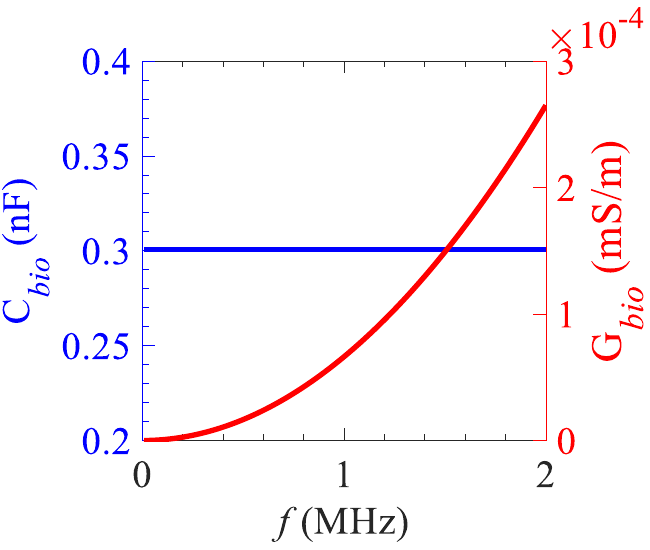}
\par\end{centering}
\caption{\label{fig:Unprtb and C}(a) A Cross section of one cell of the IDC
showing two electrodes in an unperturbed surrounding (background and
substrate). The equivalent circuit model of each domain is indicated
in this figure. (b) Capacitance and conductance values of the designed
IDC versus frequency.}
\end{figure}
\begin{equation}
C_{b}=\frac{\varepsilon_{0}\varepsilon_{rb}\pi L_{e}}{\mathrm{ln}\left(4\left(\alpha_{1}+d_{e}/w_{e}\right)\right)}\label{eq:Cb}
\end{equation}

\begin{equation}
G_{b}=\frac{\sigma_{b}L_{e}}{\alpha_{2}+\alpha_{3}\cosh^{-1}\left(1+d_{e}/w_{e}\right)},\label{eq:Gb}
\end{equation}
where $L_{e}$ and $w_{e}$ represent the length and width of each
electrode, respectively, $d_{e}$ is the gap between the two adjacent
electrodes, $\varepsilon_{rb}$ and $\sigma_{b}$ are the relative
permittivity and the conductivity of the background medium, respectively,
and $\varepsilon_{0}$ is the permittivity of vacuum. We anticipate
that the constant coefficients $\alpha_{1}=0.46$, $\alpha_{2}=0.21$
and $\alpha_{3}=0.58$ are obtained from the fitting procedure based
on the IDC parameters given hereafter for the IDC biosensor shown
in Fig. \ref{fig:BioSensor}.

Note that the equivalent capacitance and conductance of the substrate
domain are derived from (\ref{eq:Cb}) and (\ref{eq:Gb}), simply
by replacing the background material parameters with those of the
substrate, i.e., replacing $\varepsilon_{rb}$ and $\sigma_{b}$ with
$\varepsilon_{rs}$ and $\sigma_{s}$, respectively.

We assume that the inter-electrode domain is filled with the background
material and due to the small (compared to the wavelength) parallel
side walls of the adjacent electrodes, the inter-electrode domain
experiences a uniform electric field distribution, hence the equivalent
capacitance and conductance of the inter-electrode domain are estimated
by \citep{Griffiths1999}

\begin{equation}
C_{i}=\varepsilon_{0}\varepsilon_{ri}\frac{t_{i}L_{e}}{d_{e}}\label{eq:Ci}
\end{equation}
\begin{equation}
G_{i}=\sigma_{i}\frac{t_{i}L_{e}}{d_{e}}.\label{eq:Gi}
\end{equation}

Here $t_{i}$, $\varepsilon_{ri}$ and $\sigma_{i}$ are the thickness
of the electrode, the relative permittivity, and conductivity of the
inter-electrode domain, respectively. The IDC is comprised of $N$
cells repeated in the $x$ direction and the equivalent circuits of
the cells are in parallel. According to the operative frequency range
and also due to the parallel connection of the electrodes, it is observed
that the IDC shows a negligible inductance at its terminals in the
considered frequency range. With a reasonable approximation, the admittance
of a single cell of the interdigitated capacitor is approximated as

\[
Y_{\mathrm{cell}}=\underset{G_{\mathrm{cell}}}{\underbrace{(G_{b}+G_{s}+G_{i})}}+\mathrm{j}\omega\underset{C_{\mathrm{cell}}}{\underbrace{(C_{b}+C_{s}+C_{i})}.}
\]
hence, the total admittance of the biosensing capacitor is derived
as $Y_{bio}=G_{bio}+\mathrm{j}\omega C_{bio}=NY_{\mathrm{cell}}$.
Figure \ref{fig:Unprtb and C}(b) shows an example of a designed IDC
where the total capacitance is $C_{bio}=0.3\,\mathrm{n\mathrm{F}}$
and the total conductance is $G_{bio}=67\,\mu\mathrm{S}$ at $1\,\mathrm{kHz}$.
In order to meet the required design values of the time-periodic LC
sensor that exhibits an EPD, we optimize the IDC using (\ref{eq:Cb})-(\ref{eq:Gi}).
For the aforementioned design, the geometrical parameters are set
as $L_{e}=10\,\mathrm{mm}$, $w_{e}=200\,\mathrm{\mu m}$, $t_{i}=50\,\mathrm{\mu m}$,
$d_{e}=250\,\mathrm{\mu m}$, and $N=20$. We use deionized water
with $\varepsilon_{rb}=81.2$, and $\sigma_{b}=5\mathrm{\,\mu S/m}$
(at 100 MHz) as background and inter-electrode materials, and quartz
with $\varepsilon_{rs}=4$, $\sigma_{s}=0$ as substrate. The frequency
dependent behavior of the materials is considered using the Debye
model in \citep{Abeyrathne2013}.

\begin{figure}
\begin{raggedright}
(a)
\par\end{raggedright}
\noindent \begin{centering}
\includegraphics[width=3in]{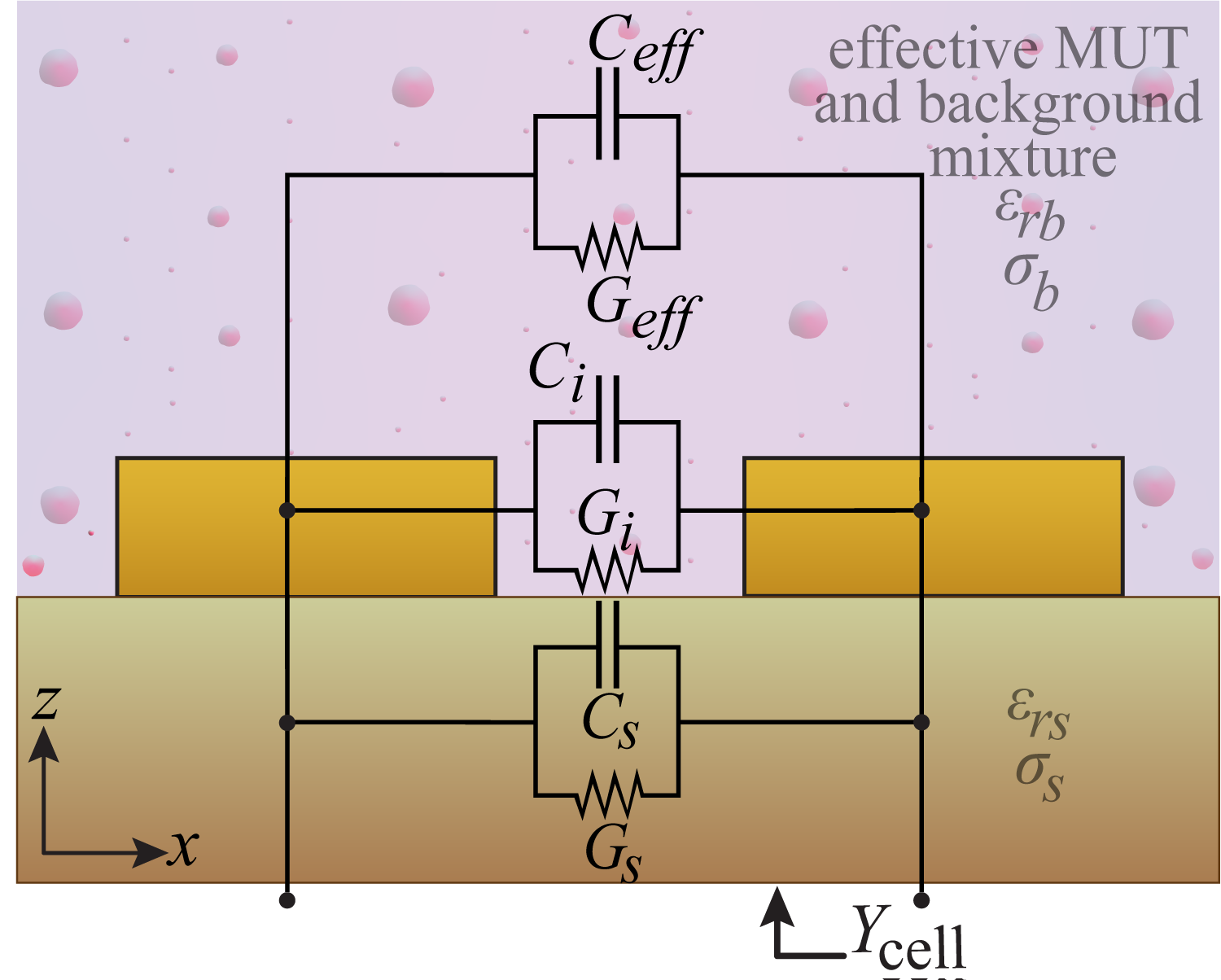}
\par\end{centering}
\begin{raggedright}
(b)
\par\end{raggedright}
\begin{centering}
\includegraphics[width=3in]{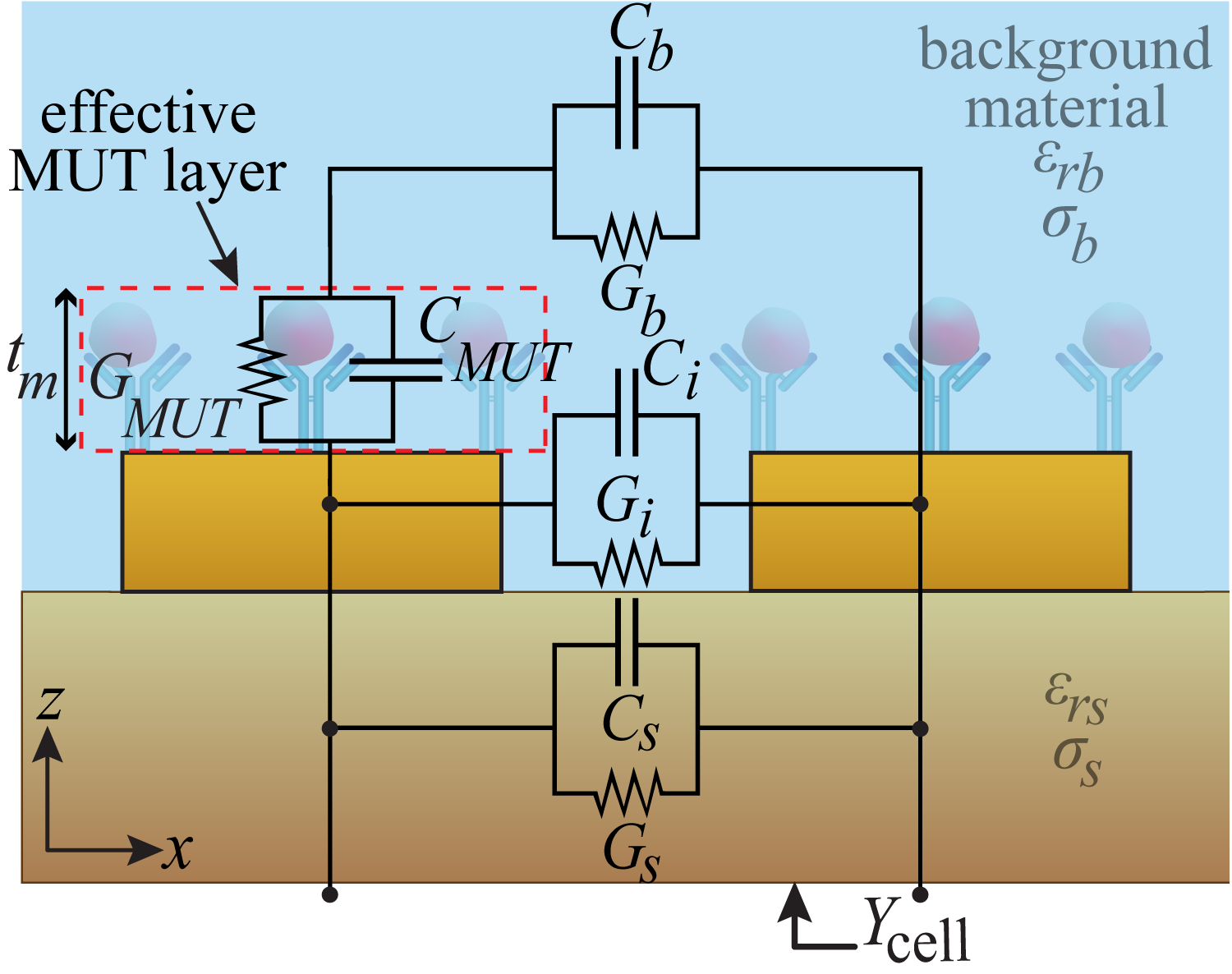}
\par\end{centering}
\begin{raggedright}
(c)~~~~~~~~~~~~~~~~~~~~~~~~~~~~~~~~~(d)
\par\end{raggedright}
\begin{centering}
\includegraphics[width=1.75in]{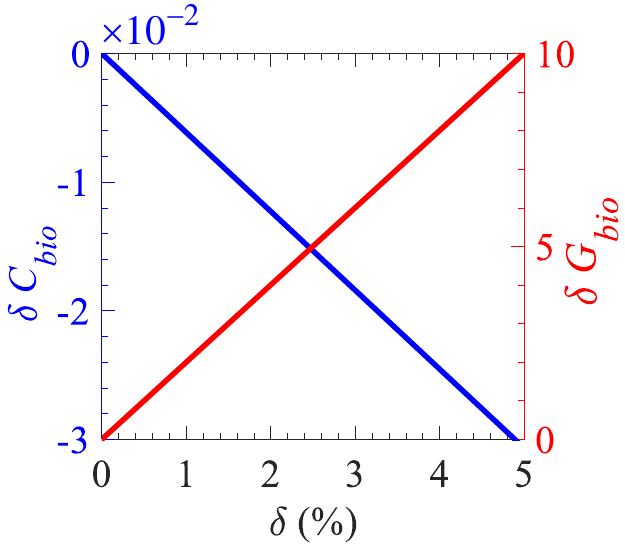}\includegraphics[width=1.75in]{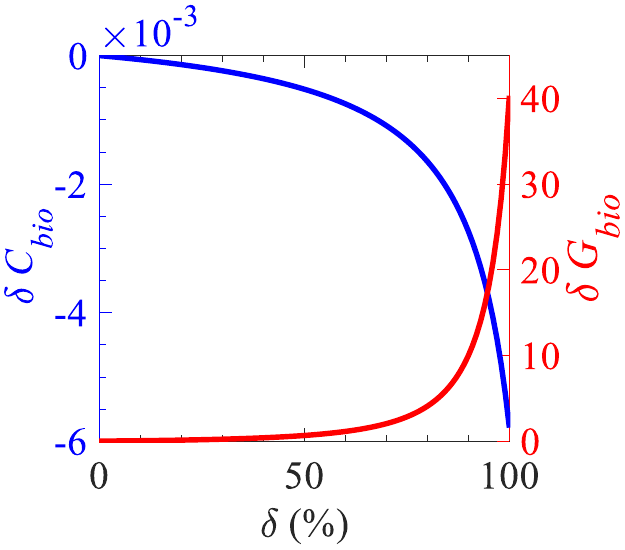}
\par\end{centering}
\caption{\label{fig:Prtb and deltaC}(a) Cross section of one cell of the IDC
where the background domain is filled with a uniformly dissolved MUT
into the background material. (b) Cross section of one cell of the
IDC where the effective layer made of the MUT and antibodies is immersed
in the background material. (c) and (d) Effect of the perturbation
$\delta$ on the equivalent circuit of the biosensor capacitor relative
to the case in (a) and (d), respectively. In (b) and (d) the thickness
of the MUT layer is only a few nanometers.}
\end{figure}

Next, we investigate the variation of the capacitance and conductance
of the IDC based on the variation of the MUT concentration by using
the derived model. According to the type of the MUT, there are two
common bionsensing modes: (i) the MUT (e.g. glucose) dissolves in
the background material uniformly changing the electromagnetic properties
of the background material, and (ii) a thin layer of the MUT (e.g.
proteins) covers the exposed surface of the electrodes. The cross
section of the perturbed biosensing capacitors for both sensing scenarios
are depicted in Fig. \ref{fig:Prtb and deltaC}(a) and (b), respectively.
Note that in an experimental scenario, the interdigitated capacitance
is typically encapsulated in a silicone chamber such as Polydimethylsiloxane
(PDMS) or Ecoflex which the temperature and humidity are kept under
control, i.e., isolating the experiment from outside \citep{dautta2020passive}.

The perturbation $\delta$ in each biosensing scenario is defined
as the relative volumetric concentration of the MUT in the background
material where $\delta=0$ represents the unperturbed structure (i.e.,
only the background material is present) and $\delta=1$ means that
the entire background material is replaced by the MUT. Such a uniform
mixture of the MUT and background material can be described by an
effective permittivity and conductivity using the linear mixture formula
for composite media \citep{Chan1986,Simpkin2010,Bruggeman1935}

\begin{eqnarray}
\varepsilon_{eff}/\varepsilon_{0} & = & \delta\varepsilon_{rm}+(1-\delta)\varepsilon_{rb}\label{eq:Ceff}\\
\sigma_{eff} & = & \delta\sigma_{m}+(1-\delta)\sigma_{b},\label{eq:Geff}
\end{eqnarray}
where $\varepsilon_{rm}$ and $\sigma_{m}$ represent the relative
permittivity and conductivity of the MUT, respectively.

To find the equivalent capacitance and conductance of the first biosensing
scenario in Fig. \ref{fig:Prtb and deltaC}(a), we substitute the
effective parameters of the composite mixture formula (\ref{eq:Ceff})
and (\ref{eq:Geff}) into (\ref{eq:Cb}) and (\ref{eq:Gb}) to derive
the equivalent circuit parameters of the perturbed biosensor capacitor
$C_{bio}$ shown in Fig. \ref{fig:Prtb and deltaC}(a). The change
in the capacitance and conductance values of the perturbed equivalent
circuit biosensor capacitor $C_{bio}$ and $G_{bio}$ are respectively
denoted by $\delta C_{bio}=(C_{bio}(\delta)-C_{bio})/C_{bio}$ and
$\delta G_{bio}=(G_{bio}(\delta)-G_{bio})/G_{bio}$ are reported in
Fig. \ref{fig:Prtb and deltaC}(c). The result in this Figure is based
on the assumption that the MUT is glucose described with $\varepsilon_{rm}=30$
and $\sigma_{m}=0.6\mathrm{\,\mu S/m}$ in the considered frequency
range \citep{7520546}.

In the second biosensing scenario, Fig. \ref{fig:Prtb and deltaC}(b),
we consider a thin MUT layer (i.e., typically a protein layer) with
relative volumetric concentration $\delta$, that adheres to the top
surface of the electrodes and is surrounded by the background material
as shown in Fig. \ref{fig:Prtb and deltaC}(b). Therefore, a layer
made of MUT and background material is formed (referred to as effective
MUT in the following) and it covers the entire electrodes' surface
through which the fringing electric field passes. Hence, the effective
layer is modeled by a circuit comprised of a capacitance $C_{MUT}$
in parallel with a conductance $G_{MUT}$, where such circuit in turn
is in series with the background domain's equivalent circuit. Since
the effective MUT layer thickness $t_{m}$ is only a few nanometers,
the electric field distribution is uniform in the this layer. The
equivalent circuit parameters of the effective MUT layer is obtained
from $C_{MUT}=\varepsilon_{eff}w_{e}L_{e}/t_{m}$ and $G_{MUT}=\sigma_{eff}w_{e}L_{e}/t_{m}$
where $\varepsilon_{eff}$ and $\sigma_{eff}$ are given by the linear
mixture formula in (\ref{eq:Ceff}) and (\ref{eq:Geff}), respectively.
Considering the equivalent circuit model of different domains represented
in Fig. \ref{fig:Prtb and deltaC}(b), we derive the variation in
the values of the perturbed equivalent circuit parameters $C_{bio}$
and $G_{bio}$ ($\delta C_{bio}$ and $\delta G_{bio}$) reported
in Fig. \ref{fig:Prtb and deltaC}(d). The result in this figure is
based on the assumption that the MUT is made of Keratin with $\varepsilon_{rm}=8$
and $\sigma_{m}=1.2\mathrm{\,\mu S/m}$, $t_{m}=50\,\mathrm{nm}$
\citep{bibi2016review}.

\section{Characterization of the biosensor performance\label{sec:Characterization-of-the}}

We characterize the sensitivity of the biosensor system operating
at an EPD, based on a LTP-LC resonator made of the parallel arrangement
of the biosensing capacitor in section \ref{sec:Analysis-of-Bio}
and a time-varying capacitor.

\subsection{Sensitivity comparison with conventional biosensors}

We start by showing a comparison between the sensitivity of a biosensor
based on a LTP-LC resonator operating near an EPD and a conventional
biosensor based on a linear time-invariant (LTI)-LC resonator, i.e.,
a standard LC resonator. The values of the LTP-LC resonator are the
same as those in the previous sections. To assess a fair comparison,
we assume that the capacitance in the LTI-LC resonator is equal to
the time average capacitance in the LTP-LC resonator, i.e., $C_{0}=(C_{1}+C_{2})/2$,
and all the other parameters are the same as those of the LTP-LC case
in Section \ref{sec:Enhance_sens}, i.e., $L_{0}=15\mathrm{\mu H}$,
$R=0.1\Omega$, $C_{bio}=0.3\,\mathrm{nF}$ and $G_{bio}=67\,\mathrm{\mu S}$.

\begin{figure}
\begin{centering}
\includegraphics[width=3.5in]{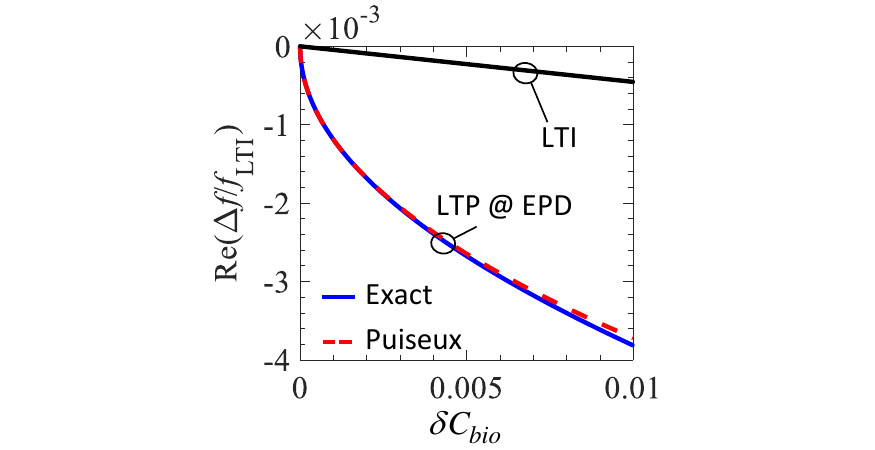}
\par\end{centering}
\caption{\label{fig:LTV_LTI} Relative change in resonance frequency (only
the one moving towards lower frequencies is shown) of a biosensor
as a function of the relative change in the biosensing capacitance
$C_{bio}$. The blue and dashed-red curves show the resonance frequency
shift of the LTP-LC biosensor with EPD, whereas the black curve shows
the resonance shift in a conventional LTI-LC resonator, respectively.
The comparison clearly shows the much higher sensitivity of the sensor
based on the EPD in the LTP-LC resonator.}
\end{figure}

Figure \ref{fig:LTV_LTI} illustrates the change in the real part
of a resonance frequency $\Delta f=f_{p}(\delta)-f_{p}(0)$ versus
relative perturbation $\delta C_{bio}$, for both the LTI (standard
case) and LTP (EPD case) biosensors. Therefore, for the LTP-LC biosensor,
$\Delta f$ describes the shift of the resonance frequency $f_{p}(\delta)$
in the perturbed biosensor with respect to the 6th harmonics ($n=6$)
of the degenerate resonance frequency, i.e., $f_{p}(0)=f_{e0}+6f_{\mathrm{m}}=(675.8+j2.6)\,\mathrm{kHz}$
of the unperturbed biosensor (i.e., working at an EPD). Similarly,
$\Delta f$ for the LTI-LC case shows the shift of the resonance frequency
with respect to unperturbed resonance frequency $f_{p}(0)=(715.4+j2.2)\,\mathrm{kHz}$.
The change $\delta C_{bio}$ in the biosensing capacitance is due
to the change in the MUT concentration $\delta$ in the background
material, shown in Figs. \ref{fig:Prtb and deltaC}(c) and (d) for
the two biosensing scenarios.

The change of the resonance frequency $\Delta f=f_{p}(\delta)-(f_{e0}+6f_{\mathrm{m}})$
based on the EPD perturbation is well described by the Puiseux series
in (\ref{eq:f_p}), truncated to the first order. Indeed this approximation
is in very good agreement with the ``exact'' result for the LTP
case obtained by solving Eq. (\ref{eq:eig_prob}), showing the analytical
nature of the ultra sensitivity concept of the EPD-based sensor. In
Fig. \ref{fig:LTV_LTI} we have normalized the perturbation of the
resonance frequency for both the LTP-LC and LTI-LC resonators to the
resonance frequency of the lossless unperturbed LTI-LC resonator that
is calculated as $f_{\mathrm{LTI}}=1/(2\pi\sqrt{L_{0}(C_{0}+C_{bio})})=715.4\,\mathrm{kHz}$.
This Figure shows the highly remarkable sensitivity associated with
the biosensor designed to operate at an EPD of the LTP-LC system.
Such ultra-sensitivity of the LTP-LC resonator operating at a second
order EPD was observed experimentally in \citep{Kazemi2019Expre}
in general terms, hence here we investigate the sensitivity to the
variation of a MUT in a biosensing scenario. A comparison with the
state of the art RF sensors such as \citep{tseng2018functional,dautta2020passive,kim2015rapid,hajiaghajani2019selective}
shows that our designed biosensor operating at an EPD has very high
sensitivity for the same MUT concentration. For instance in \citep{kim2015rapid}
the authors achieve relative resonance frequency shift of $0.03\%$
for $\delta=1\%$ whereas our biosensor shows relative resonance frequency
shift of $0.3\%$ for the same MUT concentration. One may note that
the value of $\alpha_{1}$ in Eq. (\ref{eq:f_p}) can significantly
affect the sensitivity of the LTP-LC resonator and it is determined
by the parameters of the system. Considering the values used in this
paper, $\alpha_{1}=-\mathrm{j}2.6$. Moreover, it can be inferred
from the fractional power expansion of the resonance frequencies in
Eq.(\ref{eq:f_p}) that for an imaginary value of $\alpha_{1}$, a
perturbation $\delta>0$ implies that only the real part of the resonance
frequency changes while the imaginary part is constant (i.e. there
are two purely real frequencies), whereas a perturbation $\delta<0$
implies that the imaginary part is the one changing while the real
part remains constant. Future work shall focus on how to maximize
this value based on the design of the circuit parameters.

\subsection{Sensing scalability across different frequencies}

\begin{figure}
\begin{centering}
\includegraphics[width=3.5in]{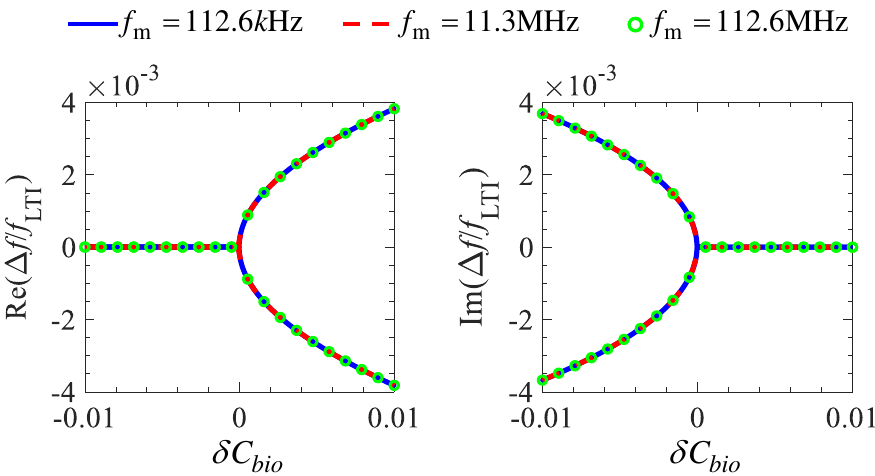}
\par\end{centering}
\caption{\label{fig:Sensitivity}Resonance frequency shift $\Delta f(\delta C_{bio})$,
normalized to modulation frequency, exhibiting large variations even
for very small relative perturbations $\delta C_{bio}$, for three
different designs with three different modulation frequencies. In
each case the LTP-LC resonator works at an EPD. This plot illustrates
the scalability of the ultra sensitivity concept over a large range
of frequencies. The solid-blue curve shows a design with modulation
frequency $f_{\mathrm{m}}=112.6\,\mathrm{k}\mathrm{Hz}$, the dashed-red
curve and green circles represent scaled designs of the biosensor
with modulation frequencies $f_{\mathrm{m}}=11.3\,\mathrm{MHz}$ and
$f_{\mathrm{m}}=112.6\,\mathrm{MHz}$, respectively.}
\end{figure}

As mentioned in the introduction, a system operating at an EPD exhibits
an enormous sensitivity to any perturbation to the system. In such
a system, shown in Fig. \ref{fig:LC_circuit}, we have different EPD
resonances for different modulation frequencies, hence we can design
a sensor with exceptional sensitivity to operate at any of them. In
order to illustrate the exotic performance of the described sensing
system, we assume that the biosensing capacitor is experiencing a
$\delta$ increase in the concentration of the MUT, hence the biosensing
capacitance value is perturbed as $C_{bio}(\delta)$, and its relative
change $\delta C_{bio}$ follows the behavior shown in Figs. \ref{fig:Prtb and deltaC}(c)
and (d), for the two sensing scenarios shown in Fig. \ref{fig:Prtb and deltaC}.
In turn, a relative positive increment $\delta C_{bio}$ perturbs
the LTP-LC resonator operating at an EPD, generating two real resonance
frequencies, whereas a negative $\delta C_{bio}$, generates two resonance
frequencies that deviate in their imaginary part, following the dispersion
diagram in Fig. \ref{fig:LC_circuit}(b). These features are better
shown in Fig. \ref{fig:Sensitivity} that illustrates such ultra-sensitivity
of the resonance frequency to small values of $\delta C_{bio}$, considering
three case with three different modulation frequencies $f_{\mathrm{m}}$.
Fig. \ref{fig:Sensitivity}(a) and (b) show the real and imaginary
shifts of the resonance frequencies, respectively, where the solid-blue
curve represents a design with a modulation frequency $f_{\mathrm{m}}=112.6\,\mathrm{k}\mathrm{Hz}$,
the dashed-red curve and green circles represent a scaled designs
of the biosensor with modulation frequencies $f_{\mathrm{m}}=11.3\,\mathrm{MHz}$
and $f_{\mathrm{m}}=112.6\,\mathrm{MHz}$, respectively.

We conclude from the figure that the real and imaginary parts of the
resonance frequency are sharply sensitive to the external perturbation
(e.g., the concentration of the MUT). We also conclude that this sensitivity
property is valid (it is actually the same) regardless of the chosen
modulation frequency which indicates a freedom in the choice of the
circuit components, and that the concepts presented in this paper
are scalable to any operating frequency.

\section{Conclusion}

We have exploited the concept of EPDs induced in linear time-periodic
systems to achieve extremely sensitive biosensors based on the detection
of a resonant frequency shift. We use a single time-varying LC resonator
whose capacitance is given by the parallel arrangement of a time-variant
capacitor and the biosensing capacitor. In our proposed
scheme with a \textit{single}
LTP resonator, the perturbation of the sensing capacitance perturbs
the system away from the EPD and results in two real-frequency shifts
from the EPD one. This is in contrast to what occurs in a two-coupled
resonator system operating at an EPD based on PT symmetry; indeed,
in a PT symmetric system, perturbing the sensing capacitor on the
lossy (sensing) side only, disqualifies the system as being PT symmetric
and as a result both of the system resonant frequencies become complex.
Furthermore, we have developed a model of interdigitated capacitors
that well describes changes of capacitance due to the variations in
the concentration of an MUT and investigated how sensitive is the
LTP-LC biosensor to such changes. We have considered two different
sensing scenarios, and an unprecedented sensitivity to the perturbations
of the time-variant LC resonator at an EPD is illustrated. The sensitivity
of the resonance frequency in a single, time-varying, LC resonator
working at an EPD to perturbations has been demonstrated to be much
higher than that of a single, time-invariant (i.e., standard), LC
resonator. The practical implementation of this sensing technology
seems straightforward since the time-modulated capacitance can be
realized with a simple multiplier controlled by a modulated voltage
pump \citep{Kazemi2019Expre} for example. The working principle for
the proposed ultra-sensitive biosensor is general and can be easily
implemented in existing systems to enhance sensitivity, paving the
way to a new class of ultra-sensitive sensors.

\section*{Acknowledgment}

This material is based upon work supported by the National Science
Foundation under Award No. ECCS-1711975.

\end{document}